\documentclass{article}
\usepackage{emulateapj}
\usepackage{apjfonts}
\usepackage{psfig}

\newcommand{\R}{{\cal R}}

\newcommand{\percent}{{\%}}

\newcommand{\iue}{{\em IUE}}
\newcommand{\hst}{{\em HST}}
\newcommand{\exosat}{{\em EXOSAT}}
\newcommand{\rosat}{{\em ROSAT}}
\newcommand{\euve}{{\em EUVE}}
\newcommand{\asca}{{\em ASCA}}
\newcommand{\rxte}{{\em RXTE}}
\newcommand{\osse}{{\em OSSE}}

\newcommand{\bllac}{1E~1415.6+2557}

\newcommand{\xspec}{{\sc xspec}}

\makeatletter
\let\@internalcite\cite
\def\cite{\def\astroncite##1##2{##1\ ##2}\@internalcite}
\def\citey{\def\astroncite##1##2{##1\ (##2)}\@internalcite}
\def\@citex[#1]#2{\if@filesw\immediate\write\@auxout{\string\citation{#2}}\fi
  \def\@citea{}\@cite{\@for\@citeb:=#2\do
    {\@citea\def\@citea{; }\@ifundefined
       {b@\@citeb}{{\bf ??}\@warning
       {Citation `\@citeb' on page \thepage \space undefined}}%
{\csname b@\@citeb\endcsname}}}{#1}}
\def\@cite#1#2{#1\if@tempswa #2\fi}
\def\@biblabel#1{}
\def\astroncite#1#2{#1\ #2}
\makeatother

\begin{document}
\slugcomment{Submitted to The Astrophysical Journal, July 5, 1999}

\lefthead{Chiang et al.}
\righthead{EUVE/ASCA/RXTE Observations of NGC~5548}

\title{Simultaneous EUVE/ASCA/RXTE Observations of NGC~5548}

\author{J. Chiang\altaffilmark{1}, 
C. S. Reynolds\altaffilmark{1,2}, O. M. Blaes\altaffilmark{3},
M. A. Nowak\altaffilmark{1}, N. Murray\altaffilmark{4},\\
G. Madejski\altaffilmark{5}, H. L. Marshall\altaffilmark{6}, 
P. Magdziarz\altaffilmark{7}}

\altaffiltext{1}{JILA, University of Colorado, Boulder CO 80309-0440}
\altaffiltext{2}{Hubble Fellow}
\altaffiltext{3}{Department of Physics, University of California, 
                 Santa Barbara CA 93106}
\altaffiltext{4}{Canadian Institute for Theoretical Astrophysics, University
                 of Toronto, Toronto, ON M5S~3H8, Canada}
\altaffiltext{5}{Laboratory for High Energy Astrophysics, NASA/GSFC, 
                 Greenbelt MD 20771}
\altaffiltext{6}{MIT Center for Space Research}
\altaffiltext{7}{Deceased 1998 August}
\setcounter{footnote}{0}

\begin{abstract}
We present simultaneous observations by \euve, \asca, and \rxte\ of
the type~1 Seyfert galaxy NGC~5548.  These data indicate that
variations in the EUV emission (at $\sim 0.2\,$keV) appear to lead
similar modulations in higher energy ($\ga 1\,$keV) X-rays by
$\sim$10--30\,ks.  This is contrary to popular models which attribute
the correlated variability of the EUV, UV and optical emission in
type~1 Seyferts to reprocessing of higher energy radiation.  This
behavior instead suggests that the variability of the optical through
EUV emission is an important driver for the variability of the harder
X-rays which are likely produced by thermal Comptonization.  We also
investigate the spectral characteristics of the fluorescent iron
K$\alpha$ line and Compton reflection emission.  In contrast to prior
measurements of these spectral features, we find that the iron
K$\alpha$ line has a relatively small equivalent width ($W_{K\alpha}
\sim 100\,$eV) and that the reflection component is consistent with a
covering factor which is significantly less than unity ($\Omega/2\pi
\sim 0.4$--0.5).  Notably, although the 2--10~keV X-ray flux varies by
$\sim \pm 25$\% and the derived reflection fraction appears to be
constant throughout our observations, the flux in the Fe~K$\alpha$
line is also constant.  This behavior is difficult to reconcile in the
context of standard Compton reflection models.
\end{abstract}


\section{Introduction}

Over the past decade, a moderately coherent picture of the broad band
continuum emission from Seyfert AGNs has emerged.  In this model, an
accretion disk around a super-massive black hole ($M \sim
10^7$--$10^8\,M_\odot$) produces thermal emission primarily in the
optical and ultraviolet; and a hot, Comptonizing corona above the disk
up-scatters these photons to produce X-rays with energies $\sim
1$--$100\,$keV.  Furthermore, the existence of strong fluorescent iron
K$\alpha$ lines at $\sim 6.4\,$keV and the so-called ``Compton
reflection humps'' above $\sim 10\,$keV in the spectra of type~1
Seyferts indicate that the disk, or some other cold, optically thick
material reprocesses the hard X-rays.

Central uncertainties in this model are the geometry of the corona and
the disk and how the radiation from one component affects the
properties and emission of the other.  A cold ($T \la {10^5}$K),
optically thick disk will absorb a large fraction of the incident
X-rays and will likely be significantly heated, thus reprocessing the
absorbed energy as additional thermal photons.  These photons will be
available for Compton up-scattering and may provide additional cooling
of the coronal electrons thereby decreasing the temperature of the
Comptonizing medium.  Hence, the size, shape, and location of the
corona relative to the disk will affect the temperature and spectral
properties of both components.

A related difficulty is the origin of the soft X-ray excess at
energies $\ga 0.1$~keV which has no plausible explanation within the
context of standard thin disk models (Koratkar \& Blaes 1999).  Any
reasonable range of disk temperatures for sub-Eddington accretion is
too low to account for this emission as an extension of the disk
thermal emission, and an extrapolation of the standard hard X-ray
thermal Comptonization spectrum down to these energies severely
underpredicts the observed flux.  However, Magdziarz et al.\ (1998)
have been able to model the soft X-ray excess {\em and} most of the
optical/UV as emission from very Thomson thick ($\tau \sim 30$), warm
(200~eV) Comptonizing clouds above the disk.  Despite the ad hoc
nature of this model, it makes the definite prediction that the
optical, UV and EUV flux should be highly correlated.

Correlated variability studies between different wavebands have been
the traditional means of determining which spectral components provide
the impetus for others.  NGC~5548, in particular, has been involved in
several multiwavelength monitoring campaigns, the best known of which
were efforts to use optical/UV spectra to characterize the Broad Line
Region via reverberation mapping (Clavel et al.\ 1991; Korista et al.\
1995).  As a by-product of these investigations, Peterson et al.\
(1991) compared the light curves of contemporaneous optical (4870\AA)
and UV (1350\AA) data from the earlier of these campaigns and showed
that essentially no lag exists between spectral bands within the
temporal resolution of the observations, $\pm 2\,$days (see also
Clavel et al.\ 1991).  If the observed optical/UV flux in type~1
Seyferts is thermal emission due to the local conversion of
gravitational energy in thin accretion disks, then the characteristic
radii from which continuum emission at these two wavelengths are
produced will be quite different.  Any correlated variability between
these two bands would then be limited by the sound crossing time in
the disk flow.  However, Courvoisier \& Clavel (1991) have shown that
the upper limits on the optical/UV lag for NGC~5548, as well as for
Seyferts NGC~4151 and Fairall~9, are several orders of magnitude
smaller than the sound crossing times.

As an explanation for this near simultaneity of the optical and UV,
Clavel et al.\ (1992; see also Malkan 1991) suggested that the
variable emission is due to reprocessing of X-rays from a source
located very near the central black hole.  This interpretation would
also be consistent with the observations of NGC~5548 by the \euve\
satellite during the 1993 \iue/\hst/ground-based BLR reverberation
mapping campaign (Marshall et al.\ 1997; Korista et al.\ 1995).
Marshall et al.\ obtained contemporaneous \euve\ data and found that
the EUV (at $\sim 0.16\,$keV) also varies with the optical and UV to
within $\la 0.25\,$days.  This result links the soft X-ray excess with
the lower wavebands and supports either a reprocessing model or one
such as that of Magdziarz et al.\ (1998).  In conflict with the former
scenario are the contemporaneous observations of the type~1 Seyfert
NGC~7469 by \rxte\ and \iue\ (Nandra et al.\ 1998).  Over a 30 day
baseline, Nandra et al.\ found that no simple relationship exists
between the variability in the X-rays and UV.  Light curves in both
wavebands exhibit large, factor of $\sim 2$ modulations on 10~day time
scales, but the peaks in the UV appear to lead the corresponding
maxima in the X-rays by $\sim 4$ days while the troughs in the two
wavebands are nearly simultaneous.  Nandra et al. suggest that this
behavior may be due to several reprocessing regions which interact at
different time scales depending on the flux state.  Previous
correlated UV/hard X-ray observations of NGC~5548 by {\em Ginga} and
\iue\ were only able to place an upper limit of $\sim 6\,$ days for
the characteristic lag between these two energy bands (Clavel et al.\
1992).  An earlier observation by \exosat\ did show an apparent lag of
$\sim 5\,$ks between the soft (0.05--2.5 keV) and hard (2--10 keV)
X-ray bands, but this was only seen during one of 12 observations,
albeit the longest one of 60~ks, and with fairly poor statistics
(Kaastra \& Barr 1989; see also Nandra et al.\ 1991).

Another aspect of the behavior of NGC~5548 which bears upon the
disk/corona geometry is the putative correlation between the derived
reflection fraction and the spectral index of the underlying hard
continuum (Magdziarz et al.\ 1998; Zdziarski, Lubi\'nski, \& Smith
1999).  Spectral analyses of X-ray data from Galactic black hole
candidates and Seyfert AGNs, including NGC~5548, show that the fitted
reflection fraction is systematically larger for softer underlying
spectra.  The natural interpretation of such a correlation is that the
reflecting medium is an important source of soft seed photons which
affect the cooling of the Comptonizing medium.  A larger covering
factor implies more soft photons, causing a lower temperature coronal
region and therefore a softer spectrum.  Since the fluorescent iron
line emission presumably also originates in this same cold, Compton
reflecting material, it provides an additional diagnostic to explore
the relationship between the covering factor and the spectral
hardness.

In order to investigate the roles of the various components in the
Seyfert disk/corona system of NGC~5548 further, we have conducted
simultaneous \euve, \asca, and \rxte\ monitoring observations.  This
combination of instruments allowed us to study the soft X-ray excess,
its relationship to the harder, Comptonized X-ray emission, and the
higher energy signatures of X-ray reprocessing in accretion disks, the
fluorescent iron K$\alpha$ line and the Compton reflection component.
We describe the observations by the various instruments in \S~2.  In
\S~3, we present an analysis of the correlated variability analysis
between different energy bands, and in \S~4 we examine the power
spectra of the various bands.  In \S~5, we analyze the X-ray spectral
properties in detail during each observation epoch.  We discuss the
implications of our observations for disk/corona models in \S~6 and
present our summary and conclusions in \S~7.

\section{The Observations and Data Reduction}
Great effort was made to ensure that the observations by the various
telescopes in this campaign were as simultaneous as possible.
Unfortunately, because of differences in time allocation, conflicts
with other scheduled observations, differing visibility constraints
and other unforeseen complications, absolute simultaneity over our
proposed 23~day baseline was not possible.  In particular, several
data gaps of a few $\times 10\,$ks exist in the \rxte\ data because of
scheduling conflicts; the \euve\ spacecraft went into ``safepoint''
mode just prior to one of our shorter monitoring observations; and the
final \rxte\ observation which took place over a $200\,$ks time period
had no contemporaneous \euve\ or \asca\ monitoring because of
visibility constraints.  Nonetheless, these data represent some of the
highest time resolution broad band monitoring of any AGN.

In order to provide uniformity in referring to the various pointings
performed by the three satellites, we have adopted a numbering scheme
based on the originally scheduled simultaneous observations.  In this
scheme, the June~15, June~18--24, July~1 and July~7 observations are
numbered 1--4, with the June~18--23 (obs.~2) subdivided as we discuss
below.  We refer to all the other observations taken by individual
instruments by the dates on which they took place.

\subsection{\euve}

Our strategy for using \euve\ to monitor the soft X-ray flux from
NGC~5548 was identical to that employed by Marshall et al.\ (1997).
Photometry was carried out using the Deep Survey (DS) detector with
the Lex/B filter, and the pointing was offset from nominal by 0.3
degrees along the direction of the SW spectrometer dispersion axis.
This was done in order to avoid the deadspot in the DS detector while
largely preserving the capabilities of the SW spectrometer.  In
extracting the count rates, we employed the analysis of Marshall et
al.\ which accounts for dead-time and ``Primbsch'' corrections as in
the standard \euve\ analysis procedure and which also applies a
vignetting correction necessitated by the offset pointing.  Another
important feature of this analysis is its implementation of a maximum
likelihood technique using the instrumental point spread function to
determine the source and background counts rather than specifying
source and background apertures.  Further information on these
procedures can be found in Marshall et al.\ (1997).

Table~\ref{euve_log} gives the details of our \euve-DS observations.
The spectral densities we report account for a Galactic column density
of $N_H \simeq 1.7\times 10^{20}$\,cm$^{-2}$ and assume a power-law
spectrum with index 0--2 ($f_\nu \propto \nu^{-\alpha}$).  For the
narrow effective bandpass we consider (70--90\AA), the resulting
spectral densities we compute are not very dependent on the value of
$\alpha$.  The 1998 June 2, June 9, and July 1 observations were
simultaneous with Lick observations scheduled on those dates, although
the June~2 Lick observation did not occur due to bad weather.  As we
noted earlier, the \euve\ spacecraft went into ``safepoint'' mode just
prior to our June~15 joint \asca/\rxte\ observations (cf.\
Tables~\ref{asca_obs} \&~\ref{RXTE_observations}), and we did not
obtain simultaneous \euve\ data for our July~7 \asca/\rxte\
observations because of a scheduling error.

\subsection{\asca}
\subsubsection{Basic data reduction}

\asca\ observed NGC~5548 on five occasions during this campaign.
The dates, good exposure times, count rates and 2--10\,keV fluxes are
reported in Table~\ref{asca_obs}.  \asca\ possesses four detectors;
two solid-state imaging spectrometers (SIS0 and SIS1) and two
gas-imaging spectrometers (GIS2 and GIS3).  These detectors are
located at the focal points of four independent but coaligned X-ray
telescopes.  During our campaign, data from all four detectors were
obtained.  SIS data taken in both {\sc bright} and {\sc faint} mode
were combined in order to maximize the total signal and a standard
{\sc grade} selection was performed in order to reduce the effects of
particle and instrumental background.  Data from the SIS were further
cleaned in order to remove the effects of hot and flickering pixels
and subjected to the following data-selection criteria: the satellite
should not be in the South Atlantic Anomaly (SAA), the object should
be at least $5^\circ$ above the Earth's limb, the object should be at
least $25^\circ$ above the day-time Earth limb, and the local
geomagnetic cut-off rigidity (COR) should be greater than 6\,GeV/$c$.
Data from the GIS were cleaned to remove the particle background and
subjected to the following data-selection criteria: the satellite
should not be in the SAA, the object should be at least $7^\circ$
above the Earth's limb and the COR should be greater than 7\,GeV/$c$.
These steps were performed using {\sc xselect v1.4} which is part of
the {\sc ftools v4.1} package.

Images, light curves and spectra were extracted from the good data
using a circular region centered on the source.  For the SIS, an
extraction radius of 3\,arcmins is used whereas a radius of 4\,arcmins
is used for the GIS.  These regions are sufficiently large to contain
all but a negligible portion of the source counts.  Background spectra
were extracted from source free regions of the same field of view
within each of the four \asca\ instruments.  Background regions
for the SIS were taken to be rectangular regions along the edges of
the source chip whereas annular regions were used to extract GIS
background.  Approximately 1\% of the photons in the source extraction
region were found to be from the background.

\subsubsection{The effects of RDD}

The effects of radiation damage to the CCD's of both SIS detectors
have reduced the ability of the on-board software to estimate and
correct for dark current.  This problem is known as residual dark
distribution (RDD) and, in these late stages of the \asca\ mission, is
becoming a major concern.  Data taken in {\sc faint} mode can have the
effects of RDD partially corrected for during the ground-based
reduction process.  Unfortunately, due to telemetry constraints, the
majority of our data were taken in {\sc bright} mode and cannot be
corrected.

Operationally, RDD renders the soft end of the SIS spectrum
($<1$\,keV) untrustworthy.  Comparing our SIS0 and SIS1 spectra
reveals major discrepancies between these two instruments below 1\,keV
which are readily attributed to RDD effects.  For this reason, we
restrict ourselves to the 2--10\,keV band when performing spectral
fitting with SIS data.

\subsubsection{An SIS/GIS discrepancy}

The spectral fitting described below was initially performed using the
2--10\,keV data from all four \asca\ instruments.  However, the GIS
spectra were found to strongly disagree with the SIS spectra at
high-energies (8--10\,keV) --- the two GIS spectra consistently lay
above the two SIS spectra in each and every observation of this
campaign.  When these spectra are fitted with a spectral model
consisting of an absorbed power-law model and a Gaussian emission line
(see below), residuals are present in the GIS spectra at energies
coincident with the Gold M-edge (2.2\,keV) and Xenon L-edge
(4.8\,keV).  This strongly indicates the use of an incorrect GIS gain
factor.  No linear gain correction could eliminate these residuals
simultaneously.  Faced with the conclusion that there is a non-linear
gain problem in our GIS data, we choose not to use GIS data in any
of the spectral fitting reported below.

Further support for the hypothesis that the GIS is in error is given
when \rxte-PCA data are fitted simultaneously with these \asca\ data.
In the overlap band (4--10\,keV), the PCA spectra generally agree well
with SIS spectra and disagree with GIS spectra.  However, to our
knowledge, the \asca\ and \rxte\ instruments have not been
cross-calibrated.  As we discuss below, we have therefore allowed the
relative normalization for the spectral models which are fit to data
from each instrument to vary, and we have also allowed the spectral
index of the hard continuum to vary separately for the SIS and PCA
data.

\subsection{\rxte}

Standard extraction methods were employed to determine the PCA and
HEXTE light curves and spectra.  The PCA instrument consists of five
collimated ($1^\circ$~FWHM) proportional counter units (PCUs) which
are numbered 0--4.  Each PCU contains three multianode detector layers
with a mixture of xenon and methane gas, has a bandpass of 2--60~keV
and a geometric collecting area of $\sim 1400\,$cm$^2$.  The energy
resolution of each detector is $\sim 8$\% FWHM at 6.6~keV (Glasser,
Odell, \& Seufert 1994).  PCA data were collected only from PCUs 0--2,
since PCUs 3 and 4 were turned on for a smaller fraction of the
on-source time due to breakdown.  This source is relatively weak, so
in order to maximize signal-to-noise, we accumulated events only from
the top xenon/methane layer of the PCA.  As recommended by the PCA
team, data were discarded for the 30 minutes following a SAA passage,
during Earth occultation (i.e., when the Earth elevation angle is $<
10^\circ$), and when there is severe electron contamination.  The
variability and spectral analyses were greatly facilitated by the
recently developed faint source ($\la 40\,$cps) background model.  It
is implemented in the {\sc pcabackest} program and is contained in the
model files {\tt pca\_bkgd\_faintl7\_e03v03.mdl} and {\tt
pca\_bkgd\_faint240\_e03v03.mdl}.  The first file contains information
on the variation on the ``L7'' rate which is a background rate made up
of a combination of 7 rates in pairs of adjacent signal anodes, while
the second file is related to the integrated recent particle doses
during SAA passages, as measured by the HEXTE particle monitor. SAA
passages and doses are recorded in {\tt
pca\_saa\_history\_v8}. Response matrices were constructed using {\sc
pcarmf}~v3.5 and {\sc pcarsp}~v2.36.  For spectral fitting, we
restricted the energy range of the PCA data from 3 to 20\,keV

The HEXTE instrument consists of two clusters of four NaI/CsI-phoswich
scintillation counters which are sensitive from 15 to 250\,keV with an
open area of $\sim 900\,$cm$^2$ for each cluster.  Beam switching or
``rocking'' of these clusters between source and background fields
provided direct measurements of the HEXTE background with an on-source
duty cycle of $\sim 60$\%.  As recommended by the HEXTE team, data
were discarded for the 2 minutes following a SAA passage and during
Earth occultation.  Standard deadtime corrections were applied to the
Science Archive data.  In order to maximize signal-to-noise in the
light curves, only counts from absolute detector channels 30--122 were
used.  As can be seen from Table~\ref{RXTE_observations}, the HEXTE
count rates were very low, $\la 1\,$cps throughout our observations.
Hence, statistically meaningful HEXTE spectra could only be
accumulated over relatively long integrations, $\ga 4 \times 10^5\,$s,
which is substantially longer than the variability time scale seen in
the PCA data.

\subsubsection{1E~1415.6+2557}
A minor complication in the \rxte\ analysis is the existence of a
contaminating source within the field-of-view of the PCA and HEXTE
collimators, $0.5^\circ$ away from NGC~5548.  The source is a
BL~Lac object, \bllac, and has been previously observed by \rosat\
(Nandra et al.\ 1993).  As an additional precaution, since Nandra et
al.\ reported several additional sources in the \rosat-PSPC FOV
surrounding NGC~5548, we performed a PCA scanning observation of the
field to determine if any other contaminants were present (cf.\
Marshall et al.\ 1998).  Using the collimator response model of the
PCA which was provided to us by the PCA instrument team, we fit the
scan observations and found that we could account for all the counts
in the field adequately with just two sources: NGC~5548 and
\bllac.

During the first four of our \rxte\ observing periods, the
contribution by \bllac\ to our NGC~5548 measurements was determined
from $\sim 3$--5\,ks pointings directed at 1E~1415.6 +2557 just prior
to and after each NGC~5548 observation period.  This bracketing was
done in case \bllac\ exhibited any variability on 30\,ks time scales.
We used the aforementioned collimator model to fit the count rates
from these two sources simultaneously during the beginning and ending
sub-portions of each observation.  We then extracted spectra from
these sub-portions for each source.  The spectra for \bllac\ were fit
with \xspec\ {\sc v10.00} using the associated background-subtracted
NGC~5548 spectrum as a ``correction file'', appropriately scaled to
yield the fitted count rates.  Table~\ref{1415.6_properties} gives the
estimated count rates and spectral fits for \bllac\ obtained from this
procedure.  When there are significant differences in the count rates
of \bllac\ for the two bracketing pointings, they are reported
separately, i.e., as 2a/b and 3a/b.

For the observations beginning 1998 August~16, we altered our strategy
for estimating the flux and spectrum from \bllac.  Rather than relying
on scaled spectral correction files extracted from direct pointings at
NGC~5548 which are in turn contaminated by flux from \bllac, we took
``background'' measurements at the symmetric location on the opposite
side of NGC~5548.  For the same roll angles, these pointings should
contain the same contribution from NGC~5548 as the direct \bllac\
pointings, but since they are $\sim 1^\circ$ away, they should be
uncontaminated by \bllac\ itself.  Correction files from these
pointings were then used in the spectral analysis.  We find that the
spectrum of \bllac\ during the August~16 observations was somewhat
harder and the count rate in the PCA had increased by $\sim 40\,$\%.
We note that this increase is not due to our altered observing
strategy as we have also applied the collimator model fitting method
and confirm this result.  The collimator response for pointings
directed at NGC~5548 to photons from \bllac\ is $\sim 0.33$, so the
contaminating contribution to the NGC~5548 light curves will be $\sim
2$\,cps with a modulation of $\la 0.8\,$cps, while the PCA count rates
for NGC~5548 are typically 20--25~cps.  Therefore, in our variability
analysis of NGC~5548, we can safely ignore contributions from \bllac.

\noindent
\parbox{0.45\textwidth}{
\centerline{\psfig{figure=fig1.ps,width=0.45\textwidth,angle=270}}
\figcaption{\euve, \asca-SIS, and \rxte-PCA light curves 
for the long observation.  All light curves have been normalized to
unity mean.  For clarity, the \asca\ and \rxte\ curves were then
offset by 0.75 and 1.5 respectively.  The dashed lines show the
boundaries of the four intervals which were used for studying the
intra-observation spectral variability. \label{compositelcfig}}}
\medskip

\section{Correlated Variability}

In Fig.~\ref{compositelcfig}, we show the \euve, \asca, and \rxte-PCA
count rate light curves for the longest of our contemporaneous
observation periods.  The most distinct feature in all three of these
light curves is the large step at $\sim 2\times 10^4$\,s in the
figure.  This step is more pronounced ($\sim 40$\%) at EUV energies
than in the harder X-rays ($\sim 20$\% for the \rxte-PCA data).  After
this step, in the\rxte\ light curve there are several local maxima
which recur on $\sim 50\,$ks time scales; similar modulations also
appear in the \euve\ light curve, but with less certainty due to the
relatively poor statistics.  As a first order analysis of the spectral
variability across these wavebands, we have computed cross-correlation
functions (CCFs) for each pair-wise combination.

The method we use is the Z-transformed Discrete Cross-Correlation
Function (ZDCF) of Alexander (1997) based upon the DCF method of
Edelson \& Krolik (1988).  In the latter procedure, which was designed
to accommodate unevenly and differently sampled data trains, pair-wise
combinations of measured flux values from the two light curves are
binned according to their relative delays or lags.  For the data pairs
within each lag bin, a quantity which is essentially Pearson's linear
correlation coefficient is computed (Press et al.\ 1992).  In the
continuum limit, this procedure is equivalent to the standard
definition of the CCF.  Alexander (1997) proposed two modifications of
this method: first, the lag bins are defined to contain a specific
number of data pairs which should be $> 11$ in order to ensure
convergence; second, the errors are estimated by applying the
Z-transform to the correlation function values and using the number of
pairs for the given bin and known properties of the Z-transform to
estimate the errors.  For identical binning, the ZDCF estimates of the
CCF are equal to those produced by the DCF, but the uncertainties are
better behaved.  Auto-correlation functions can also be computed by
this method.

Figure~\ref{ccfs} shows the ZDCFs for \euve\ vs 0.5--1\,keV \asca-SIS,
\euve\ vs \rxte-PCA, and 0.5-1\,keV \asca-SIS vs \rxte-PCA.  In each
case, a positive time delay indicates the latter light curve lagging
the former.  In order to obtain a quantitative estimate of the
relative lags, we fit the ZDCF values in the vicinity of the peak with
a parabola (solid curves in Fig.~\ref{ccfs}).  This functional form
has no specific physical meaning---it is merely a 
\parbox{0.45\textwidth}{
\centerline{\epsfysize=0.55\textwidth\epsfbox{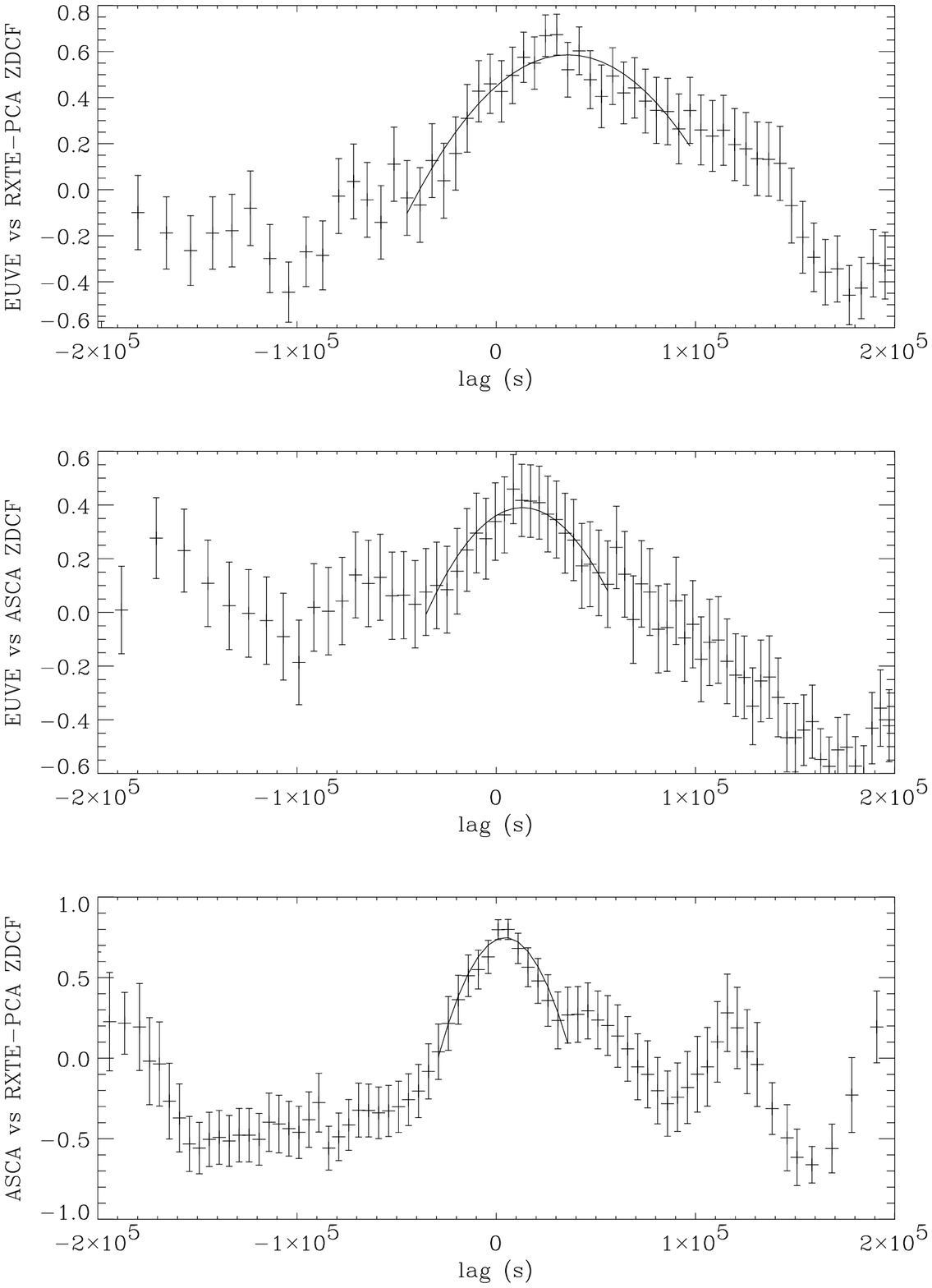}}
\figcaption[fig2.eps]{Cross-correlation functions computed using 
the ZDCF (Alexander 1997).  The bands which are compared are \euve\
(0.14--0.18\,keV), \asca\ (0.5--1\,keV), and \rxte-PCA (2--20\,keV)
(cf.\ Fig.~\protect{\ref{compositelcfig}}).  The solid curves are the
parabolas which were fit to find the location of the characteristic
lags. \label{ccfs}}}
\bigskip

\noindent
convenient parametrization of the ZDCF shape near the maximum.  From
these fits, we see that higher energy light curve modulations are
delayed relative to lower energy ones with the variations in the
2--20\,keV \rxte-PCA data lagging those of the \euve\ by $\sim 35$\,ks
and those of the 0.5--1\,keV \asca\ data by $\sim 5$\,ks.  Similarly,
the 0.5--1\,keV \asca\ data lag the \euve\ by $\sim 13$\,ks.  We
determine confidence limits using Monte Carlo methods where for each
trial we produce simulated light curves taking the measured light
curves as input and assuming Gaussian statistics.  In
Table~\ref{lags}, we report the mean value of the fitted peaks and the
99.9\% C.L.s for each pair of light curves.  We have also estimated
the cross-correlation function using the Lomb-Scargle method for
computing the FFT of unevenly sampled data (Scargle 1989) and find
essentially the same results.

\section{Power Spectra}

We can combine our pointed X-ray observations, which span time scales
from approximately $5\times10^3$\,s to $3\times10^5$\,s, with
observations from the All Sky Monitor (ASM) on board the \rxte\
spacecraft, which span time scales ranging from days to 3 years.  We
have used these data to construct an X-ray power spectrum that covers
four orders of magnitude in Fourier frequency from
$10^{-8}$--$10^{-4}$\,Hz. The result is presented in
Fig.~\ref{fig:psd}.  A power spectrum that covers a similarly broad
range of time scales, but with much better statistics, has been
obtained from pointed \rxte\ observations of the Seyfert 1 galaxy
NGC~3516 (\cite{edelson:99a}).

As the orbital time scales of the \rxte, \asca, and \euve\ spacecraft
are approximately 5\,ks, significant variability power (due to SAA
passages and source occultation) appears on these time scales and at
higher harmonics.  In addition, for the \euve\ 
\parbox{0.45\textwidth}{
\centerline{\epsfysize=0.45\textwidth\epsfbox{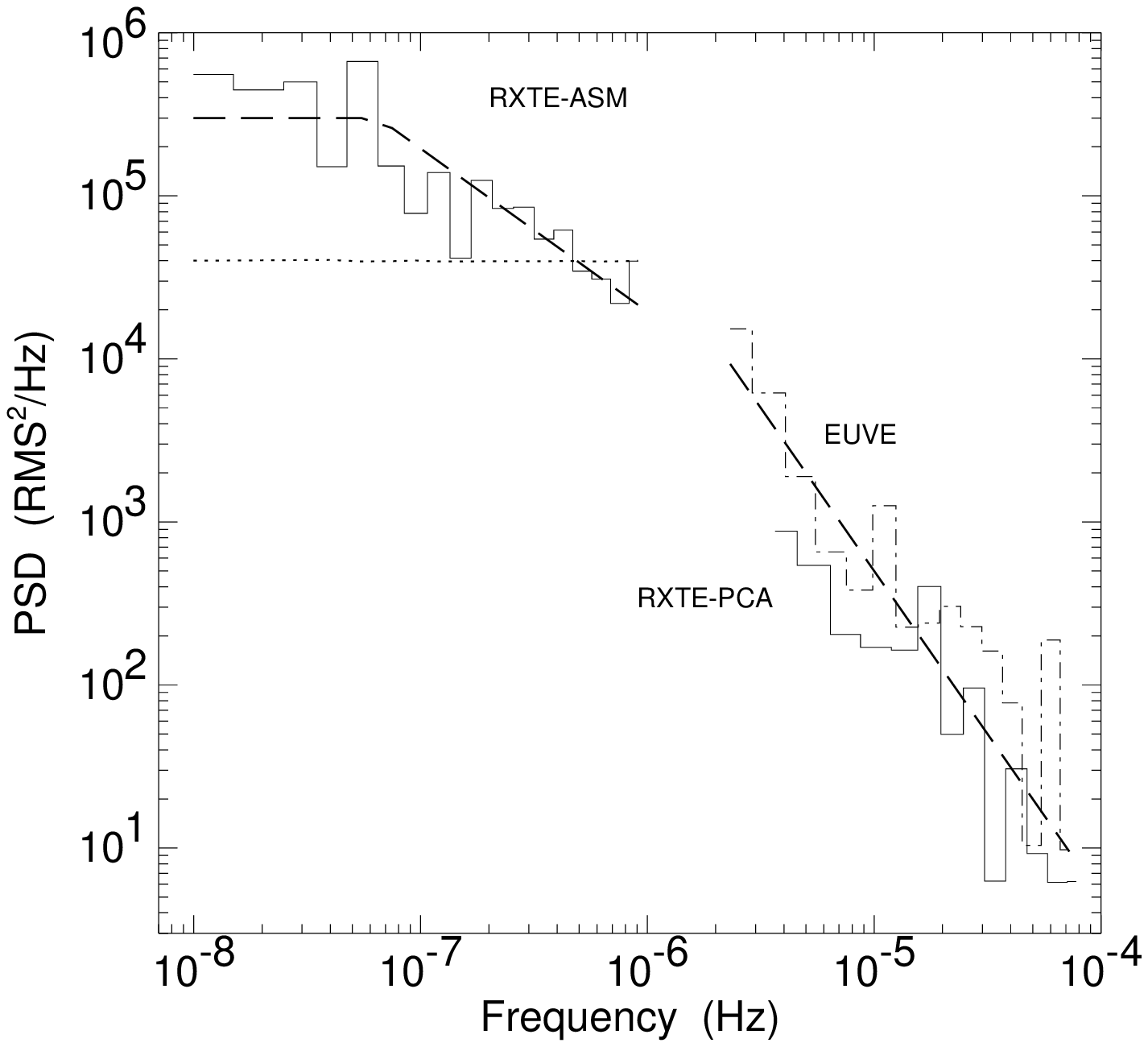}}
\figcaption[fig3.eps]{Power Spectral Densities constructed from 
\rxte-ASM,  \rxte-PCA, and \euve\ data (see text).  Dashed lines show 
flat, $f^{-1}$, and $f^{-2}$ overlaid on the power spectra.  The
dotted line show our estimate of the mean noise level of the ASM data.
All PSDs are normalized such that integrating over positive Fourier
frequencies yields the root mean square variability, normalized to the
mean, for each particular light curve.  \label{fig:psd}}}
\bigskip

\noindent
spacecraft, we must integrate for approximately 5\,ks in order to
obtain good signal to noise.  We therefore only consider Fourier
frequencies $< 10^{-4}$\,Hz.  The PSDs are constructed from light
curves with 5.6\,ks time bins for the \euve\ data, and 128\,s time
bins for the \rxte-PCA data.  These light curves, however, still
contain a number of data gaps, therefore we use techniques for
calculating power spectra of unevenly sampled light curves (see
\cite{lomb:76a,scargle:82a}).  The PSDs are then normalized so that
integrating over positive frequencies yields the root mean square
(rms) variability relative to the mean (see
\cite{belloni:90a,miyamoto:92a}).  The resulting PSDs are then
logarithmically binned over frequencies ranging from $f\rightarrow
1.2f$.  For the \euve\ PSD we calculate that the noise contribution to
the rms variability is $11.0\pm0.6\%$ of the mean, therefore we
subtract this (assumed white) noise level from the \euve\ PSD.  The
noise level of the \rxte\ PSD is negligible.

Several results are immediately apparent in Fig.~\ref{fig:psd}.
First, we see that the \euve\ light curve is more highly variable than
the \rxte-PCA light curve, with an rms variability of $18.1\pm1.4\%$
compared to $7.4\pm0.6\%$ over the time scales sampled by the PSDs.
Both the \rxte-PCA and the \euve\ PSDs are approximately $\propto
f^{-2}$.  The $f^{-2}$ PSD seen for the \rxte-PCA light curve is
comparable, in both shape and rms amplitude, to the high-frequency ($>
3$\,Hz) PSDs seen for galactic black hole candidates (GBHCs), such as
Cygnus~X-1, observed in their low/hard states
(\cite{miyamoto:92a,nowak:99a}).

The low-frequency X-ray PSD is calculated via ASM data.  The ASM
provides light curves in three energy bands, 1.3--3.0\,keV,
3.0--5.0\,keV, and 5.0--12.2\,keV, typically consisting of several
90\,s measurements per day (see
\cite{levine:96a,remillard:97a,lochner:97a}).  We have combined all
three energy channels into a single channel, and then rebinned the
data on 5 day time scales.  Here we have filled data gaps (39 out of
233 points) with a linear interpolation, and the power spectrum is
then constructed using standard FFT techniques.  We did not subtract
the expected noise level calculated from the ASM error bars, as the
derived PSD does not flatten at this level.  This indicates that
either the ASM error bars are overestimated or that they do not follow
gaussian statistics.

The derived PSD is approximately flat at low frequencies, shows a
possible break at $\approx 7\times 10^{-7}$\,Hz ($\approx 200$~day
period), and is consistent with an $f^{-1}$ dependence at higher
frequencies.  The extent to which the PSD remains flat at low
frequencies is uncertain due to the limited time-span of the ASM
observations compared to the putative 200 day break period.  We note
that Czerny, Schwarzenberg-Czerny, \& Loska (1999) find a break in the
optical power spectrum at a period $\la 1000$\,days.  The slope of the
ASM-PSD at high frequencies is also uncertain due to the large noise
level (shown as a dotted line in Fig.~\ref{fig:psd}).  The rms
amplitude of the ASM lightcurve, including noise, is $\approx 30\%$.
A Lomb-Scargle periodogram yields comparable results, and also shows a
formally significant (via the methods of \cite{horne:86a}) peak at a
200 day period.  This peak is coincident with the break in the PSD,
and furthermore the methods of Horne \& Bailunas are only strictly
valid for a single sinusoidal period buried within pure counting
noise.

The overall X-ray PSD is again remarkably similar in both shape and
rms amplitude to those seen in the low/hard states of GBHCs.  Low
state GBHCs such as GX~339$-$4 (\cite{nowak:99c}) or Cyg~X-1
(\cite{belloni:90a,miyamoto:92a,nowak:99a}) typically have an rms
amplitude of 30--40\%, a flat PSD at low frequency with a break into
an $f^{-1}$ spectrum at $0.03$--$0.3$\,Hz, and a further break into an
$f^{-2}$ spectrum at 1--10\,Hz.  The X-ray PSD we have found for
NGC~5548 is comparable to this, except scaled to approximately a
factor of $10^6$ lower in frequency (although we do not resolve the
break into the $f^{-2}$ spectrum).  A similar result was found for the
X-ray PSD of NGC~3516 (\cite{edelson:99a}).

The factor of $10^6$ is comparable to the expected ratio of the black
hole masses between these two kinds of systems.  Furthermore, the
$10^4$\,s time delays between the \asca\ and hard \rxte\ bands
discussed above are scaled by a similar factor compared to the
$10^{-3}$--$10^{-2}$\,s X-ray time lags seen at the 1--10\,Hz break
frequency in the low/hard state of GBHCs (see
\cite{miyamoto:89a,nowak:99a}).  In the high-frequency regime in
GBHCs, the variability coherence between X-ray bands begins to break
down (\cite{vaughan:97a,nowak:99a,nowak:99c}).  In these regimes, we
therefore do not expect a perfect linear correlation between energy
bands.  Carrying the analogy to GBHCs further, low/hard state GBHCs
exhibit X-ray time lags of 0.03--0.1\,s at the lower break frequency
(\cite{miyamoto:89a,nowak:99a,nowak:99c}), hence one might expect
${\cal O}(10^5~{\rm s})$ time delays in the observed ${\cal
O}(100~{\rm day})$ time scale variability.  This is likely below the
detection threshold of the \rxte-ASM, and we find no evidence for such
delays comparing the lowest and highest ASM channels.  Such time
delays, however, if they were to exist, will likely be detectable by
future all sky monitors with larger effective area and somewhat longer
sampling time scales.

\section{X-ray spectral fitting}

In this section we use standard spectral fitting techniques to
characterize the X-ray energy spectrum of NGC~5548.  From previous
X-ray studies, we expect the spectrum to be dominated by non-thermal
power-law continuum emission.  In addition, a relativistically broad
iron emission line (from the surface of the accretion disk) and an
associated reflection continuum will be superposed on this power-law
continuum.  These expectations guide our choice of spectral models
(described below), and are confirmed in the sense that our spectral
models describe the data very well.  As we noted previously, we
restrict our attention to data above 2\,keV because of RDD effects;
this removes the necessity to model the warm absorber in this object.
All spectral fitting has been performed using {\sc xspec v10.00}.

In order to search for spectral variability, we examined X-ray spectra
from ten distinct time intervals.  For seven of these intervals, we
have simultaneous \asca/\rxte\ data: observations 1, 3 and 4, as well
as four sub-intervals of observation 2 (defined in
Fig.~\ref{compositelcfig}).  The three additional intervals for which
only \rxte\ observed occurred just prior to and after observation~3
and during the period 16--18~August~1998
(cf.~Tables~\ref{RXTE_observations} and~\ref{spectral_fits}).  This
set of spectra allows us to probe both long (weekly) and short (daily)
time scale spectral variability.  The overlapping \asca\ and \rxte\
data were pruned in order to ensure strict simultaneity of the spectra
from the two instruments.  This was necessary since the observations
for both satellites were interrupted either due to scheduling
conflicts (\rxte) or telemetry failures (\asca).  However, since we
are not examining spectra from data which are integrated over
intervals of less than several tens of ks, we have neglected
non-simultaneity on shorter time scales such as the orbital period.

Since the count rates obtained from the \rxte-PCA observations are
much greater than that of the \asca-SIS, we consider the \rxte-PCA
data by itself initially.  We then use those fits to guide our joint
SIS/PCA fits, relying upon the greater spectral resolution of the
\asca\ detectors to constrain the iron line component better.

\subsection{The \rxte\ data}

We fit several models to these data.  Our initial model is a simple
power-law modified by Galactic neutral absorption ($N_{\rm
H}=1.7\times 10^{20}\,{\rm cm}^{-2}$; $z=0$).\footnote{For all the
fits to the \rxte-PCA data, we include a fixed power-law to represent
the contamination by \bllac.  The spectral indices used for this added
power-law are shown in Table~\ref{1415.6_properties}.  The flux
normalizations at 1\,keV shown in that table are reduced by a factor
determined by the collimator response to \bllac\ for each pointing
directed at NGC~5548. This factor is typically $\sim 0.33$.} This
proves to be an unacceptable fit in all ten cases, with residuals
clearly indicating the need for an iron emission line and spectral
hardening above 10\,keV.  Next, we add a narrow emission line to the
power-law spectrum with a (rest-frame) energy fixed to that of the
K$\alpha$ line of cold iron (6.4\,keV).  Using the F-test, this leads
to a significant improvement (at greater than the 99\% level) in the
goodness-of-fit in all of the datasets.  Adding more complexity to the
fit, the iron line is allowed to have a broad Gaussian profile and
arbitrary energy.  Again using the F-test, this leads to a significant
improvement (at the 90\% level) in the goodness-of-fit in two cases
only (observations 1 and 2.2).

Proceeding to more physical models, we replace the Gaussian emission
line profile with the iron line profile expected from a relativistic
accretion disk around a Schwarzschild black hole (Fabian et al.\
1989).  The inner edge of the line emitting region, the inclination of
the disk, and the normalization of the line are free parameters in
this fit.  Hidden fixed parameters are the outer edge of the line
emitting region, $r_{\rm out}$, the line emissivity index $\beta$
defined such that the local line emissivity is proportional to
$r^{-\beta}$, and the rest frame line energy which is fixed at
6.4\,keV.  The emissivity index $\beta$ is fixed at $\beta=2.5$ which
approximates the expected emissivity in a variety of physical
scenarios.  The outer radius $r_{\rm out}$ is fixed at $1000r_{\rm
g}$, where $r_{\rm g}=GM/c^2$ is the gravitational radius of the
central black hole.  For our chosen value of $\beta$, the fits are
insensitive to the value of $r_{\rm out}$, provided that $r_{\rm
out}>100r_{\rm g}$.  Although this is a more physical model, the
\rxte-PCA data cannot distinguish it from the Gaussian line profile,
i.e., there is no improvement in the goodness-of-fit in any of the
datasets when the diskline profile is used.  We do note, however, that
both the broad Gaussian and the diskline fits are consistent with a
substantial fraction of the iron emission originating from within
$\sim 20\,r_g$.

Our final models include the Compton reflection continuum which is
implemented in \xspec\ as the {\sc pexrav} model (Magdziarz \&
Zdziarski 1995).  We fix the $e$-folding energy of the underlying hard
continuum at 120\,keV as determined from fits by Magdziarz et al.\
(1998) to data from \osse\ observations of NGC~5548, and we set the
inclination of the planar cold reflector at $30^\circ$.  We also fix
the chemical abundances to be Solar.  The relative normalization of
this component, $\R$, is left as a free parameter.  For an isotropic
primary X-ray source, $\R=\Omega/2\pi$, where $\Omega$ is the solid
angle subtended by the cold reflecting medium as seen by an observer
situated at the X-ray source.  This leads to a large improvement in
the goodness-of-fit for all of the observations, showing that the
Compton continuum (which produces the high-energy spectral hardening)
is clearly detected.

\subsection{Joint \asca/\rxte\ spectral fitting}

Four of our \asca\ observations were scheduled to be contemporaneous
with \rxte\ observations.  Joint spectral fitting of simultaneous data
from \asca-SIS and \rxte-PCA can be a powerful technique --- it allows
us to utilize both the spectral resolution of \asca\ and the
broad-band sensitivity of \rxte.  The results of the joint
\asca-SIS/\rxte-PCA fits for our seven intervals are listed in
Table~\ref{spectral_fits}.  To facilitate direct comparison, we show
the fits to the \rxte-PCA data alone which include a broad Gaussian
emission line.  The addition of the \asca\ data does not affect the
model fits greatly, although in some cases it allows the iron line
width to be resolved.

\subsection{Observed correlations between spectral parameters}

Here we use the results of our spectral fitting of the \asca\ and
\rxte\ data to examine spectral variability.  In order to ensure that
we have a set spectral parameters which have been uniformly analyzed,
we consider only those parameters which have been obtained from fits
to the \rxte\ data alone.  Spectral parameters from the joint
\asca/\rxte\ fits yield similar results.  Figure~\ref{fluxcorrfig}
shows the photon index $\Gamma$, the iron line flux $N_l$, the iron
line equivalent width $W_{K\alpha}$, and relative reflection
normalization $\R$ as a function of the 2--10\,keV flux, $F_{\rm
2-10}$, of the hard continuum component.  For these plots and all of
the fits to the spectral trends discussed in this section, we present
uncertainties as 1-$\sigma$ error bars.

\subsubsection{The photon index}

There is a clear positive correlation between the
photon index and the 2--10\,keV flux (Fig.~\ref{fluxcorrfig}a).
Fitting a simple linear form to this correlation gives
\begin{equation}
\Gamma=(1.63\pm 0.04)\,+\,(0.030\pm 0.006)\left(\frac{F_{\rm
    2-10}}{10^{-11}\,{\rm erg}\,{\rm s}^{-1}\,{\rm cm}^{-2}}\right).
\end{equation}
This simple model does not strictly provide a formally adequate
description of the spectral variation with flux, giving a
goodness-of-fit of $\chi^2/{\rm dof}=16.1/8$.  However, we note that
most of the contribution to this large value of $\chi^2$ is due to the
Aug 16--18 data point.  Omitting this point, we find $\chi^2/{\rm dof}
= 5.7/7$ with the fit parameters being substantially unchanged.
Nonetheless, the fact that the X-ray continuum gets softer when
brighter is consistent with the trend observed in previous monitoring
campaigns of this source (Magdziarz et al.\ 1998).  However, here we
appear to find a somewhat weaker dependence of spectral index on flux.

\subsubsection{Iron line and reflection features}

Correlations of the iron line strength with the continuum flux are of
great importance.  In the simple X-ray reflection scenario, we would
expect $W_{K\alpha}$ to be constant provided that the light crossing
time of the fluorescing region is much smaller than the time scale on
which spectral variability is being probed.  If the light crossing
time of the emission line region is much greater than the time scale
being probed, a constant iron line flux, $N_l$, would be expected.
When these two time scales are comparable, reverberation effects come
into play and both $N_l$ and $W_{K\alpha}$ will be seen to vary.

Like $W_{K\alpha}$, the parameter $\R$ should measure the amount of
X-ray reflection relative to the direct continuum.  Indeed,
$W_{K\alpha}$ should be proportional to $\R$ provided the following
conditions are satisfied:
\begin{enumerate}
\item The Compton reflection continuum is not starting to dominate the
  observed continuum at the iron line energies.  In practice, this
  condition implies $\R<2$--3.
\item The ionization state of the illuminated regions of the reflecting
  medium is fixed.
\item The primary continuum has a fixed energy spectrum.  Of course,
  we have just shown that this is not strictly the case and that the
  photon index of the primary continuum emission changes by
  $\Delta\Gamma \sim 0.2$ during our campaign in a manner that is well
  correlated with the 2--10\,keV flux.  However, the Monte Carlo
  simulations of George \& Fabian (1991) show that such small photon
  index changes only have a small (less than 10\%) effect on the iron
  line equivalent width.
\end{enumerate}
Apart from inclination dependences (which are usually weak) and
ionization gradients, the geometry of either the reflecting medium or
the X-ray source does not affect the expected proportionality of
$\R$ and $W_{K\alpha}$.

\begin{figure*}[t]
\centerline{\epsfxsize=0.8\textwidth\epsfbox{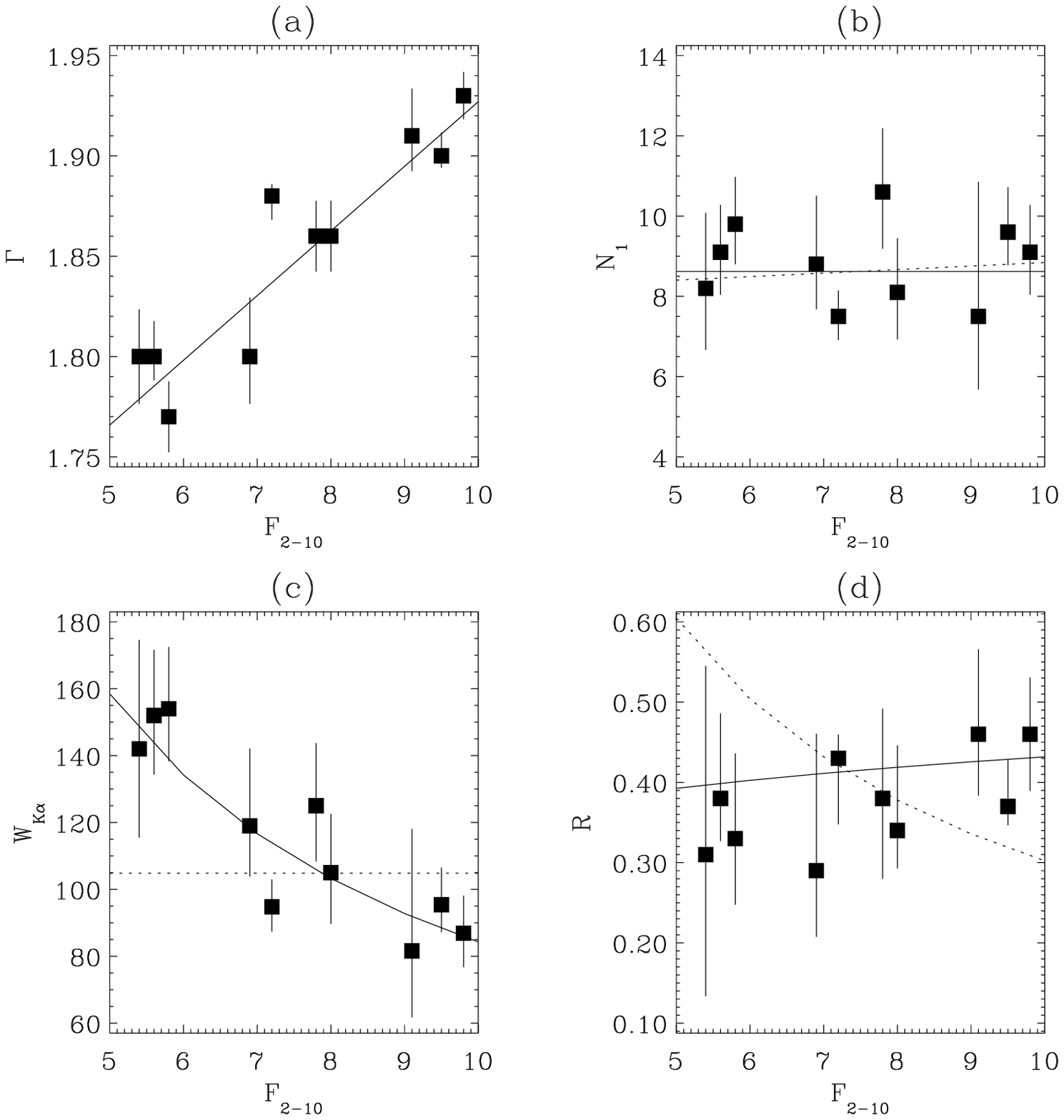}}
\figcaption[fig4.eps]{(a) Photon index $\Gamma$, (b) iron line flux 
$N_l$, (c) iron line equivalent width $W_{K\alpha}$, and (d) relative
reflection normalization $\R$ as functions of the 2--10\,keV hard
continuum flux for our ten time intervals.  The iron line flux is in
units of $10^{-5}\,$ergs\,cm$^{-2}$\,s$^{-1}$, and the 2--10\,keV hard
continuum flux is in units of $10^{-11}\,{\rm ergs}\,{\rm
s}^{-1}\,{\rm cm}^{-2}$.  The dotted curves are the expected
relationships in simple reflection models, while the solid curves are
the best fits assuming various functional forms (see text).  One-sigma
error bars are shown. \label{fluxcorrfig}}
\end{figure*}

Our spectral results hint at a more complex picture.  Inspection of
Fig.~\ref{fluxcorrfig}b shows that the iron line flux is essentially
constant.  This is borne out by the linear (dotted line) and constant
(solid) models which we have fit and which describe these data very
well.  Apart from effects introduced by the subtle changes in photon
index, this implies that $W_{K\alpha}$ is inversely proportional to
the flux (Fig.~\ref{fluxcorrfig}c).  More formally, fitting the 10
data points in the $W_{K\alpha}$-$F_{\rm 2-10}$ plot with a constant
model results in $\chi^2/{\rm dof}=20.8/9$ (dotted curve),
unacceptable at the 97\% level.  Assuming, instead, a power-law
relationship between these two parameters such that
$W_{K\alpha}\propto F_{2-10}^\alpha$ results in a dramatic improvement
in the goodness-of-fit with $\chi^2/{\rm dof}=7.9/8$ and
$\alpha=-0.9\pm 0.4$ (solid curve in Fig.~\ref{fluxcorrfig}c).  Thus,
these results are consistent with a constant line flux and an
equivalent width which is inversely proportional to the continuum
flux.

\bigskip

Since the relative normalization of the Compton reflection continuum
$\R$ should be proportional to $W_{K\alpha}$ under a fairly general
set of conditions, we would expect that $\R\propto F_{\rm 2-10}^{-1}$.
Fitting this relationship to the data depicted in
Fig.~\ref{fluxcorrfig}c gives $\chi^2/{\rm dof}=14.5/9$ (dotted
curve).  This is formally inconsistent with the data at the
91\percent\ level.  Using a more general power-law form $\R\propto
F_{\rm 2-10}^\alpha$ leads to a vastly improved fit of $\chi^2/{\rm
dof}=3.4/8$ with $\alpha=0.14\pm 0.60$ (solid curve).  This is
consistent with a constant relative reflection normalization and
implies that $W_{K\alpha}$ and $\R$ do not exhibit the proportionality
which is normally expected.  In order to examine this explicitly, we
plot $W_{K\alpha}$ vs $\R$ in Fig.~\ref{W_vs_R}.  The dotted line is
the linear relationship expected for a cold reflector with solar
abundances for which $W_{K\alpha} = 150\,$eV when $\R = 1$ (George \&
Fabian 1991).  This model yields a $\chi^2/{\rm dof} = 73/10$ and is
definitely in conflict with the data.  However, if we fit for the
proportionality constant, we obtain $W_{K\alpha} = 259\,$eV at $\R=1$
and $\chi^2/{\rm dof} = 14.4/9$ which is still formally inconsistent
with the data at the 89\percent\ level.

\section{Implications for Disk/Corona Models}

If we assume that the \euve\ light curve is representative of the seed
photons for the higher energy X-rays, we can make a simple estimate of
the size of the scattering region based upon the delays seen.  With
each scattering, a seed photon picks up a fractional energy $\Delta
E/E \simeq 4k_bT/m_e c^2$, where $T$ is the temperature of the
electrons in the corona.  The energy of a Comptonized
photon will be
\begin{equation}
E \simeq E_0 \left(1 + \frac{4k_bT}{m_e c^2}\right)^n,
\end{equation}
where $E_0$ is the initial seed photon energy and $n$ is the number of
scatterings (Rybicki \& Lightman 1979).  The time delay of this photon
with respect to the seed photon source will be roughly proportional to
the number of scatterings, $t \simeq n t_0$.  Here $t_0 \sim l_T/c$
where $l_T$ is the mean free path for Thomson scattering (or the size
of the corona if it is optically thin).  We take the effective photon
energy in the 0.5--1\,keV \asca\ data to be 
\parbox{0.45\textwidth}{
\centerline{\epsfxsize=0.5\textwidth\epsfbox{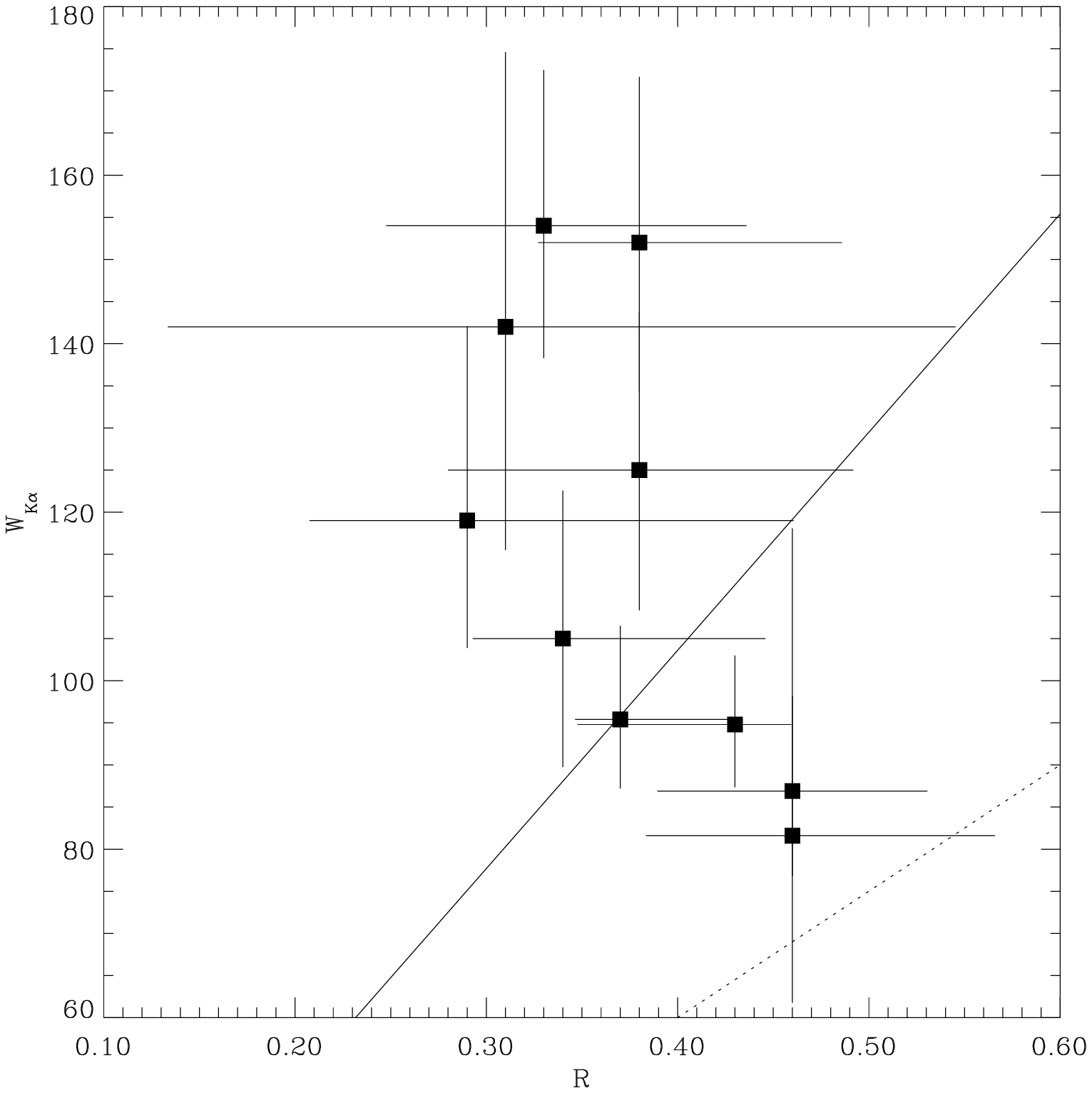}}
\figcaption[fig5.eps]{Iron line equivalent width $W_{K\alpha}$ versus 
the relative Compton reflection normalization $\R$.  The dotted line
is the expected linear relationship assuming solar abundances for the
cold reflector (George \& Fabian 1991).  The solid line is the
best-fit proportionality relationship which is still in conflict with
the data at the 89\percent\ level. \label{W_vs_R}}}
\bigskip

\noindent
$E_a = 0.78$\,keV and the effective photon energy of the 2--20\,keV
\rxte-PCA data to be $E_{\rm pca} = 5.4\,$keV.  Both of these energies
are determined using the instrument sensitivities in the relevant
bands and assuming a photon spectral index of 1.9 and a neutral
absorbing column of $1.7 \times 10^{20}\,$cm$^{2}$.  Using a value of
$k_bT \sim 50\,$keV which is found from fits to the \osse\ spectrum of
NGC~5548 (Magdziarz et al.\ 1998) and the inferred lags of the
\rxte-PCA and 0.5--1\,keV \asca-SIS light curves relative to that of
the \euve, we solve for the characteristic time between scatterings
and find $t_0 \sim 3.8\,$ksec.\footnote{One can also use the direct
lag measured between the
\asca-SIS and \rxte-PCA light curves of $5\,$ks.  From that lag, one
finds $t_0 \sim 0.9\,$ks.}  For a homogeneous corona with Thomson
depth of $\tau \sim 2$ this implies a coronal size scale of $\sim 2
\times 10^{14}$\,cm, which is of order $10 r_g$ for a $10^8\,M_\odot$
black hole.

This interpretation for the relationship between the EUV and higher
energy X-ray flux is not without its complications.  First of all, as
we note above, the fractional size of the step in the EUV is larger
than the steps in the higher energy bands.  Second, the amplitude of
the \euve\ PSD differs from that of the \rxte-PCA
(Fig.~\ref{fig:psd}).  Therefore a strictly linear relationship
between the fluxes in these bands is not possible.  One possibility is
that the \euve\ observations are sampling the Wien tail of a
black-body spectrum so that the observed variations are more
pronounced and are not necessarily characteristic of the bulk of the
seed photons.  A more provocative explanation would be that the
increase in soft photon flux produces excess cooling in the corona,
changing some combination of the corona temperature, size, or optical
depth.  The softening of the underlying continuum with increased flux
(Fig.~\ref{fluxcorrfig}a) does in fact show that properties of the
corona must be changing.

The iron line fluxes and reflection fractions also have significant
implications for the corona/disk geometry.  As we noted above, both
the reflection fraction and the iron line equivalent width are
substantially smaller than has been previously seen for this object as
well as for other type~1 Seyferts (e.g., MCG$-$6-30-15, Lee et
al.~1998; and IC~4329A, Madejski et al.~1995), for which values of $\R
\sim 1$ and $W_{K\alpha} \ga 160$\,eV are more typical.  The lower
values we find for NGC~5548 would be consistent with a substantially
smaller covering factor for the cold material.  This could occur, for
example, if the Comptonizing region lay inside an inner disk radius as
has been posited for models of Cygnus~X-1 (Dove et al.\ 1997;
Gierli\'nski et al.\ 1997).  Alternatively, the disk could extend
beneath the corona but may be completely ionized preventing this
material from contributing to either the line fluorescence or the
reflection hump.  Mildly relativistic motion of the Comptonizing
medium away from the disk could also account for the lower reflection
fractions and iron line equivalent widths (Reynolds \& Fabian 1997;
Beloborodov 1999).

However, a lack of proportionality between the iron line equivalent
width and reflection fraction (Fig.~\ref{W_vs_R}) cannot be accounted
for simply by the relative geometry of the disk and corona.  A
constant iron line flux would suggest that a substantial portion of
that emission is produced very far from the Comptonizing X-ray source,
perhaps in the outer disk or obscuring torus, so that the line
emission does not vary on time scales as short as that of the
underlying continuum.  We would then expect that the reflection hump
is likewise produced in this same distant material so that its
absolute normalization should also be constant.  Unfortunately, there
is the additional complication that the Gaussian and diskline fits to
the fluorescent iron line indicate that the line is broad and
redshifted.  The line shapes are consistent with a significant
fraction of this emission originating from an accretion disk within
$\sim 20\,r_g$.  Furthermore, spectral fits to the \asca-SIS data
which include a narrow line at 6.4\,keV along with the diskline model
yield a narrow component that can contribute only about 15\% to the
total line flux.  Therefore, based on the spectral shapes of the iron
line, a significant amount of reprocessing by distant cold material is
unlikely.

Wherever the cold reprocessor is located, these results appear to be
somewhat in conflict with models in which the spectral variability of
the underlying continuum is linked to the relative reflection
normalization $\R$ (Magdziarz et al.\ 1998; Zdziarski et al.\ 1999).
In Fig.~\ref{R_vs_Gamma}, we plot the relative Compton reflection
normalization versus the photon spectral index of the underlying
power-law (filled squares, solid contours).  As we note above, these
data are formally consistent with a constant value of $\R$, but there
does seem to be a slight linear trend in the same sense predicted by
the models.  For direct comparison, we also plot in
Fig.~\ref{R_vs_Gamma} the spectral parameters found by Zdziarski et
al.\ (1999; see also Magdziarz et al.\ 1998) which are derived from
{\em Ginga} observations.  In this figure, we show the 1-$\sigma$
error contours for both sets of data.  The spectral parameters we find
appear to occupy a different region of the $\R$-$\Gamma$ plane than
that found by these previous observations.  However, consideration of
systematic uncertainties may alter this conclusion.  {\em Ginga}-LAC
observations of the Crab have yielded a photon index of $\Gamma = 2.08
\pm 0.03$ (Turner et al.\ 1989) which agrees with early rocket
measurements by Toor \& Seward (1974).  In contrast, power-law fits to
\rxte-PCA observations of the Crab yield a significantly softer index
of $\Gamma = 2.187$ (Wilms et al.\ 1999).  If this difference of
$\Delta\Gamma \sim 0.1$ is due to a systematic shift in measured
power-law spectral index, then one may be tempted to shift each of our
values of $\Gamma$ downward by this amount, making 
\parbox{0.45\textwidth}{
\centerline{\epsfxsize=0.5\textwidth\epsfbox{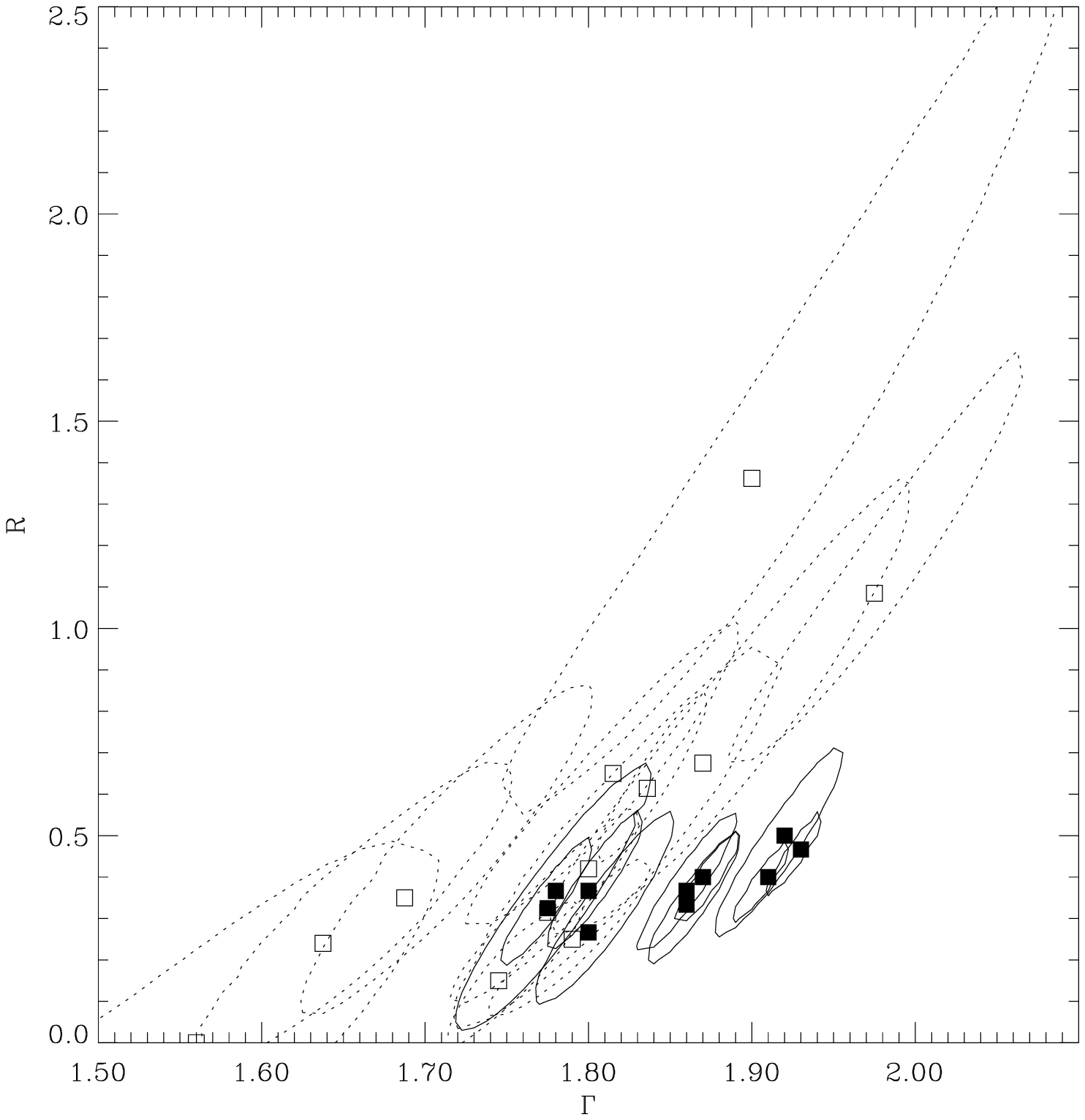}}
\figcaption[fig6.eps]{The relative Compton reflection normalization 
$\R$ versus photon spectral index $\Gamma$ for our fits to the
\rxte-PCA data (filled squares, solid error contours) and for fits to
{\em Ginga} data as presented in Zdziarski et al.\ (1999; unfilled
squares, dotted contours).  Although the \rxte-PCA parameters are
formally consistent with a constant value of $\R$, a linear trend may
be present.  However, the dependence on $\Gamma$ found for these data
is weaker that that found by Zdziarski et al.\ (1999), and these
parameters appear to occupy a different part of the $\R$-$\Gamma$
plane. One-sigma contours are shown. \label{R_vs_Gamma}}}
\bigskip

\noindent
the two sets of parameters consistent.  Unfortunately, the orientation
of the error contours clearly indicates a correlation between fitted
values of $\R$ and $\Gamma$.  This seems to be an intrinsic feature of
the model.  If so, any systematic shift of $\Gamma$ probably also
implies a bias in the fitted value of $\R$ as well.  It is therefore
not entirely clear whether these two datasets can be made consistent
even if one considers the possible systematic errors.

As a final caveat, we note that the strength of the reflection
component can be masked by features in the hard continuum which need
not be well-represented by a power-law.  Stern et al.\ (1995) point
out that for Comptonizing regions which are localized and which are
illuminated anisotropically the resulting hard continuum will have
spectral breaks just above the peak energies of the twice scattered
photons (see also Haardt 1993).  For sufficiently high values of the
compactness ($\equiv L_{\rm diss}\sigma_T/h m_e c^3$ where $L_{\rm
diss}$ is the power heating the corona and $h$ is the coronal size
scale), these breaks may occur at energies such that the reflection
component will be underestimated.  In addition, as the compactness
increases, the spectrum will soften and the break will move to lower
energies.  Hence, any variability in the absolute normalization of the
reflection component which is inferred from {\sc pexrav}-type spectral
fits could be significantly affected by changes in the compactness.
This sort of uncertainty can only be resolved by more sophisticated
spectral models.
\vspace{12pt}

\section{Summary and Conclusions}

Here we summarize the main results of this work:
\begin{enumerate}
\item The EUV leads the harder X-rays, rejecting scenarios in which 
      the soft X-ray component is produced by reprocessing of harder
      X-rays.  In this respect, these hard X-ray lags are reminiscent
      of the X-ray lags seen in NGC~7469 (Nandra et al.\ 1998).

\item However, in contrast to NGC~7469, the shorter lags we
      find for NGC~5548 are consistent with Compton diffusion time
      scales for a relatively small corona of size $\sim 10^{14}$\,cm.

\item The power spectrum of the X-ray variability shows a break at
      $\sim 200$\,day time scales.  The PSD is similar to that seen in
      NGC~3516; and if one assumes a $10^8\,M_\odot$ black hole for
      NGC~5548, this scales appropriately with mass compared to PSDs
      seen for the low/hard states of GBHCs such as Cyg~X-1 or
      GX~339$-4$.

\item The iron line equivalent width and relative reflection 
      normalizations are smaller than those which are typically
      observed for type 1 Seyferts.

\item The broad, redshifted iron line profiles, the variability of 
      the iron line equivalent width, and the apparent constancy of
      the relative reflection normalization are difficult to reconcile
      in the context of simple reflection models.

\end{enumerate}

The last point, while not definitively constrained by the data, may
pose a serious challenge for models of reflection by cold material
and will only be resolvable with further detailed temporal and
spectral observations.  More sensitive studies with future instruments
such as {\it XMM} would be extremely useful to verify this result.

\acknowledgements
We would like to thank Andrzej Zdziarski for useful conversations.
JC, MAN, and NM thank the Program on Black Hole Astrophysics at the
Institute for Theoretical Physics (ITP) at the University of
California, Santa Barbara for hosting them during part of this effort.
The ITP is supported by NSF Grant PHY 94-07194.  JC was supported by
the NASA/ADP grant NAG5-7897 and the NASA/RXTE grant DPR S-92675F.
Part of this work was performed while JC held a National Research
Council-NRL Research Associateship.  CSR thanks support from NASA
under LTSA grant NAG5-6337 and support from Hubble Fellowship grant
HF-01113.01-98A awarded by the Space Telescope Institute, which is
operated by the Association of Universities for Research in Astronomy,
Inc., for NASA under contract NAS 5-26555.  OB was supported by
NASA/RXTE grant NAG5-7125.  We would like to thank the \rxte\ Guest
Observer Facility for help with the data analysis and Tod Strohmeyer
for providing software for constructing the PCA collimator model.  In
conducting this research, we have made extensive use of the NASA
Astrophysics Data System (ADS) Abstract and Article services.  We
acknowledge use of the ADS Digital Library; the NASA/IPAC
Extragalactic Database (NED) which is operated by the Jet Propulsion
Laboratory, Caltech, under contract with NASA; and the HEASARC data
archive.

\begin{table}[p]
\centering
\begin{tabular}{lcccc}
\hline\hline
Obs. & Date & Exp. time (ks) & mean count rate ($10^{-2}$\,cps)
& Spectral Density at 76\AA ($\mu$Jy)\\
\hline
  & 02 Jun & ~~8.8 & $18.0 \pm 0.6$ & 345\\
  & 09 Jun & ~11.5 & $~8.4 \pm 0.4$ & 161\\
2 & 18 Jun & 139.8 & $~9.3 \pm 0.1$ & 178\\
3 & 01 Jul & ~21.3 & $~2.1 \pm 0.2$ & ~40\\
\hline
\end{tabular}
\caption{\euve-DS observations}
\label{euve_log}
\end{table}

\begin{table}[p]
\begin{tabular}{lccccccc}
\hline\hline
Obs. & Date & good SIS exp. & good GIS exp. & SIS0 count & GIS2 count &
     2--10 keV flux \\
     &      & time (ks) & time (ks) & rate (cps) & rate (cps) & 
     ($10^{-11}\,{\rm erg}\,{\rm cm}^{-2}\,{\rm s}^{-1}$)\\
\hline
1 & 15 Jun 1998& 21.2 & 25.3 & 2.61 & 1.56 & 5.9 \\
2 & 20 Jun 1998& 111.1 & 106.3 & 3.07 & 1.71 & 6.9 \\
3 & 01 Jul 1998& 9.2 & 10.6 & 1.66 & 0.97 &  4.2 \\
4 & 07 Jul 1998& 14.7 & 16.9 & 1.73 & 1.08 & 4.3 \\
  & 19 Jan 1999& 8.2 & 9.0 & 1.84 & 1.09 & 4.7 \\
\hline
\end{tabular}
\caption{Basic parameters of our \asca\ observations of NGC~5548.  The quoted
  fluxes are those observed (i.e. subject to the total line-of-sight
  absorption).   A simple power-law model was used to measure these fluxes.}
\label{asca_obs}
\end{table}

\begin{table}[p]
\centering
\begin{tabular}{llcccc}
\hline\hline
Obs. & Date(s) & target & Good Time (ks) & PCA/PCUs 0--2 & HEXTE rate \\
     &         &        &                & rate (cps)    & (cps) \\
\hline
1 & Jun~15--16    & NGC~5548 & ~21.24 & 25.9 & 1.0 \\
  &               & \bllac   & ~~8.82 & ~4.8 & $\cdots$ \\
2 & Jun~19--24    & NGC~5548 & 110.28 & 28.7 & 0.7 \\
  &               & \bllac   & ~~7.56 & ~4.4 & $\cdots$ \\
  &          & NGC 5548 scan & ~~3.48 & $\cdots$ & $\cdots$ \\
3 & Jun~29--Jul~2 & NGC 5548 & ~41.70 & 19.3 & 0.6\\
  &               & \bllac   & ~~7.44 & ~5.2 & $\cdots$ \\
4 & Jul~7         & NGC 5548 & ~24.60 & 19.3 & 1.0 \\
  &               & \bllac   & ~~3.60 & ~4.2 & $\cdots$ \\
  & Aug~16--18    & NGC 5548 & ~44.92 & 23.8 & 0.3\\
  &               & \bllac   & ~~7.74 & ~7.3 & $\cdots$ \\
\hline
\end{tabular}
\caption{\rxte\ observations}
\label{RXTE_observations}
\end{table}

\begin{table}[p]
\centering
\begin{tabular}{lccc}
\hline\hline
Instruments & $t_{\rm lag}$ (ks) & \multicolumn{2}{c}{99.9\% C.L. (ks)} \\
\hline
\euve\ vs \asca\     & ~13.1 & $-3.9$ & ~29.5 \\
\euve\ vs \rxte\ PCA & ~35.3 & ~23.1 & ~53.1 \\
\asca\ vs \rxte\ PCA & ~~5.0 & ~~1.9 & ~~8.0 \\
\hline
\end{tabular}
\caption{Lags obtained using the ZDCF for the three pair-wise 
combinations of instruments during the 18--23~June 1998 observation
period.}
\label{lags}
\end{table}

\begin{table}[p]
\centering
\begin{tabular}{lccc}
\hline\hline
Obs. & PCA/PCUs 0--2 rate (cps) & $\Gamma$ & $A$ ($10^{-3}$) \\
\hline
1  & $4.8 \pm 0.2$ & $2.3 \pm 0.1$ & $ 9.8 (-1.3/+3.8)$ \\
2a & $3.9 \pm 0.3$ & $2.4 \pm 0.2$ & $11.6 (-2.8/+3.8)$ \\
2b & $4.8 \pm 0.3$ & $2.2 \pm 0.2$ & $ 8.8 (-2.1/+2.8)$ \\
3a & $5.4 \pm 0.2$ & $2.3 \pm 0.1$ & $11.1 (-2.0/+2.4)$ \\
3b & $4.9 \pm 0.2$ & $2.2 \pm 0.1$ & $ 9.3 (-1.6/+2.0)$ \\
4  & $4.2 \pm 0.3$ & $2.3 \pm 0.1$ & $11.6 (-2.3/+2.9)$ \\
Aug 16--18 & $7.3 \pm 0.2$ & $2.1 \pm 0.1$ & $10.8 (-1.0/+1.0)$ \\
\hline
\end{tabular}
\caption{Properties of \bllac.}
\label{1415.6_properties}
\end{table}

\small
\begin{table}[p]
\centering
\begin{tabular}{lcccccccccccc}
\hline\hline
Obs. & $F_{\rm 2-10}$ & \multicolumn{2}{c}{$\Gamma$} & $\R$ &  
     \multicolumn{3}{c}{\sc Gaussian} & $\chi^2/{\rm dof}$ \\
     &                &      SIS    &      PCA       &      &
     $E$ (keV) & $\sigma$ (keV) & $N_l$  & \\
\hline
1   & 8.0 & $\cdots$ & $1.86\pm 0.03$ & $0.34^{+0.18}_{-0.08}$ & 
     $6.15^{+0.10}_{-0.11}$ & $< 0.41$ & $8.1^{+2.3}_{-2.0}$ & 47.4/40 \\
    & 7.9 & $1.89\pm 0.03$ & $1.86^{+0.03}_{-0.01}$ & $0.40^{+0.20}_{-0.07}$ &
     $6.15^{+0.13}_{-0.12}$ & $0.35^{+0.17}_{-0.11}$ & $7.0^{+1.7}_{-1.5}$ & 
     382/381 \\
2.1 & 7.8 & $\cdots$ & $1.86\pm 0.03$ & $0.38^{+0.19}_{-0.17}$ & 
     $6.15^{+0.11}_{-0.12}$ & $0.33^{+0.16}_{-0.18}$ & $10.6^{+2.7}_{-2.4}$ &
     38.4/40 \\
    & 7.7 & $1.88^{+0.05}_{-0.04}$ & $1.86\pm 0.03$ & $0.40^{+0.19}_{-0.18}$ &
     $6.17^{+0.10}_{-0.11}$ & $0.30^{+0.18}_{-0.15}$ & $8.1^{+1.9}_{-1.7}$ &
     327/324 \\
2.2 & 9.8 & $\cdots$ & $1.93\pm 0.02$ & $0.46\pm 0.12$ &
     $6.10\pm 0.12$ & $0.44^{+0.13}_{-0.12}$ & $9.1^{+2.0}_{-1.8}$ & 36.0/40 \\
    & 9.7 & $1.94^{+0.03}_{-0.02}$ & $1.91\pm 0.02$ & $0.41^{+0.16}_{-0.10}$ &
     $6.11\pm 0.12$ & $0.35^{+0.26}_{-0.09}$ & $6.5^{+2.5}_{-0.9}$ & 501/409 \\
2.3 & 9.5 & $\cdots$ & $1.90^{+0.02}_{-0.01}$ & $0.37^{+0.10}_{-0.04}$ &
     $6.18^{+0.13}_{-0.07}$ & $0.35^{+0.15}_{-0.11}$ & $9.6^{+1.9}_{-1.4}$ &
     29.8/40 \\
    & 9.4 & $1.93\pm 0.02$ & $1.90^{+0.02}_{-0.01}$ & $0.40^{+0.14}_{-0.05}$ &
     $6.17^{+0.15}_{-0.06}$ & $0.41^{+0.12}_{-0.10}$ & $8.1\pm 1.3$ & 423/416 \\
2.4 & 9.1 & $\cdots$ & $1.91^{+0.04}_{-0.03}$ & $0.46^{+0.18}_{-0.13}$ &
     $6.32^{+0.22}_{-0.29}$ & $< 0.74$ & $7.5^{+5.7}_{-3.1}$ & 32.2/40 \\
    & 8.8 & $1.95^{+0.06}_{-0.04}$ & $1.91^{+0.05}_{-0.03}$ & 
     $0.49^{+0.30}_{-0.12}$ & $6.25^{+0.25}_{-0.20}$ & $0.46^{+0.30}_{-0.23}$ &
     $6.2^{+3.4}_{-1.9}$ & 241/286 \\
Jun 29 & 6.9 & $\cdots$ & $1.80^{+0.05}_{-0.04}$ & $0.29^{+0.29}_{-0.14}$ &
     $6.21\pm 0.12$ & $< 0.36$& $8.8^{+2.9}_{-1.9}$ & 26.0/40 \\
3   & 5.6 & $\cdots$ & $1.80^{+0.03}_{-0.02}$ & $0.38^{+0.18}_{-0.09}$ &
     $6.20^{+0.09}_{-0.82}$ & $0.38^{+0.14}_{-0.13}$ & $9.1^{+2.0}_{-1.8}$ &
     28.0/40 \\
    & 5.3 & $1.78^{+0.06}_{-0.04}$ & $1.82^{+0.04}_{-0.02}$ &
     $0.62^{+0.33}_{-0.13}$ &  $6.16\pm 0.14$ & $0.44^{+0.17}_{-0.11}$ &
     $7.3\pm 1.4$ & 202/278 \\
Jul 2 & 5.4 & $\cdots$ & $1.80\pm 0.04$ & $0.31^{+0.40}_{-0.30}$ &
     $6.21^{+0.18}_{-0.14}$ & $< 1.11$ & $8.2^{+3.2}_{-2.6}$ & 26.3/40\\
4   & 5.8 & $\cdots$ & $1.77\pm 0.03$ & $0.33^{+0.18}_{-0.14}$ & 
     $6.17^{+0.09}_{-0.10}$ & $0.39^{+0.12}_{-0.10}$ & $9.8^{+2.0}_{-1.7}$ &
     47.0/40 \\
    & 5.7 & $1.77^{+0.05}_{-0.01}$ & $1.75^{+0.03}_{-0.01}$ &
     $0.32^{+0.20}_{-0.07}$ & $6.20^{+0.07}_{-0.10}$ & $0.29^{+0.14}_{-0.12}$ &
     $7.0^{+0.9}_{-1.0}$ & 287/332 \\
Aug 16--18 & 7.2 & $\cdots$ & $1.88^{+0.01}_{-0.02}$ & $0.43^{+0.05}_{-0.14}$ &
     $6.08^{+0.11}_{-0.04}$ & $< 0.17$ & $7.5^{+1.1}_{-1.0}$ & 29.1/40\\
\hline
\hline
Obs. & $F_{\rm 2-10}$ & \multicolumn{2}{c}{$\Gamma$} & $\R$ &
     \multicolumn{3}{c}{\sc diskline} & $\chi^2/{\rm dof}$ \\
     &                &     SIS     &      PCA       &      &
     $r_{\rm in}$ & $i$ (deg) & $N_l$ & \\
\hline
1   & 7.9 & $1.89\pm 0.03$ & $1.86\pm 0.03 $ & $0.39^{+0.19}_{-0.18}$ &
     $9.5^{+7.2}_{-3.5}$ & $< 32$ & $6.8^{+1.9}_{-1.5}$ & 383/381 \\
2.1 & 7.7 & $1.88\pm 0.04$ & $1.85\pm 0.03 $ & $0.39^{+0.19}_{-0.17}$ &
     $10.3^{+6.2}_{-3.5}$ & $< 28$ & $8.3^{+1.8}_{-1.7}$ & 327/324 \\
2.2 & 9.6 & $1.95\pm 0.02$ & $1.91\pm 0.02 $ & $0.42^{+0.10}_{-0.12}$ &
     $7.6^{+3.7}_{-1.6}$ & $< 29$ & $6.8^{+1.5}_{-1.4}$ & 503/409 \\
2.3 & 9.4 & $1.93\pm 0.02$ & $1.90\pm 0.02$ & $0.38^{+0.12}_{-0.11}$ &
     $8.9^{+6.7}_{-2.9}$ & $26^{+6.6}_{-8.2}$ & $8.1^{+1.6}_{-1.5}$ & 420/416 \\
2.4 & 8.8 & $1.96\pm 0.05$ & $1.91\pm 0.04$ & $0.50^{+0.28}_{-0.23}$ &
     $12.1^{+77}_{-6.1}$ & $41^{+18}_{-20}$ & $6.1^{+2.0}_{-2.3}$ & 242/286 \\
3   & 5.4 & $1.78^{+0.06}_{-0.04}$ & $1.82^{+0.03}_{-0.04}$ &
     $0.59^{+0.27}_{-0.25}$ & $< 16.4$ & $< 37$ & $7.3^{+1.6}_{-1.7}$ &
     203/278\\
4   & 5.7 & $1.77\pm 0.04$ & $1.75\pm 0.03$ & $0.32^{+0.18}_{-0.16}$ &
     $11.3^{+7.0}_{-3.3}$ & $< 25$ & $6.9^{+1.6}_{-1.3}$ & 288/332 \\
\hline
\end{tabular}
\caption{The parameters obtained from spectral fits to the \rxte\ and 
\asca\ data.  The spectral models consist of Galactic absorption, a
fluorescent iron emission line, an underlying cut-off power-law and a
Compton reflection component.  For fits in the upper part of the
table, a broad Gaussian function is used to model the iron line; in
the lower part, a diskline model is used.  For the joint
\asca-SIS/\rxte-PCA fits, we report separate spectral indices for the
underlying power-law for SIS and PCA.  The 2--10\,keV continuum flux
is reported for the PCA fits only and is in units of
$10^{-11}\,$ergs\,cm$^{-2}$s$^{-1}$.  The iron line flux, $N_l$, is in
units of $10^{-5}\,$ergs\,cm$^{-2}$s$^{-1}$, and the inner radius of
the diskline model, $r_{\rm in}$ is in units of $r_g = GM/c^2$.}
\label{spectral_fits}
\end{table}

\normalsize

\clearpage

%\clearpage
%
%\setcounter{figure}{0}
%
%\begin{figure}
%\epsfxsize=\hsize
%\centerline{\epsfbox{fig5.eps}}
%\caption{}
%\end{figure}
%
%\begin{figure}
%\epsfxsize=\hsize
%\centerline{\epsfbox{fig6.eps}}
%\caption{}
%\end{figure}

\end{document}